\documentclass[aps,prl,twocolumn,showpacs,groupedaddress]{revtex4-1}
\usepackage{graphicx}
\usepackage{graphics}
\usepackage{epsfig}
\usepackage{amsmath}
\usepackage{amssymb}
\usepackage{bbm}
\usepackage{verbatim}
\usepackage{psfrag}
\newcommand{\Eqref}[1]{Eq.~(\ref{#1})}
\newcommand{\Figref}[1]{Fig.~\ref{#1}}

\newcommand{\s}[1]{\sigma}

\def\ket#1{\left|#1\right>}

\begin{document}

\title{Optical Superradiance from Nuclear Spin Environment of Single Photon Emitters}
\author{E. M. Kessler$^1$, S. Yelin$^{2,3}$, M. D. Lukin$^{3,4}$, J. I. Cirac$^1$, and G. Giedke$^1$}
\affiliation{$1$ Max--Planck--Institut f\"{u}r Quantenoptik,
Hans-Kopfermann--Str. 1, D--85748 Garching, Germany\\
$2$ Department of Physics, University of Connecticut, Storrs, CT 06269, USA\\
$3$ ITAMP, Harvard-Smithsonian Center for Astrophysics, Cambridge, MA 02138, USA\\
$4$ Department of Physics, Harvard University, Cambridge, MA 02138, USA}
\date{\today}

\begin{abstract}
  We show that superradiant optical emission can be observed from the
  polarized nuclear spin ensemble surrounding a single photon emitter such as a single
  quantum dot (QD) or Nitrogen-Vacancy (NV) center. The superradiant light is
  emitted under optical pumping conditions and would be observable
with realistic experimental parameters.
\end{abstract}
\maketitle

Superradiance (SR) is a cooperative radiation effect resulting from
spontaneous build-up and reinforcement of correlations between initially
independent dipoles. Its most prominent feature is an emission
intensity burst in which the system radiates much faster than an otherwise
identical system of independent emitters. This phenomenon is of fundamental
importance in quantum optics and since its first prediction by Dicke in 1954
\cite{Dicke1954} it has been studied extensively (for a review see
\cite{Gross1982}). The rich steady state properties of the associated dynamics
can account for strong correlation effects including phase transitions and
bistability \cite{Carmichael1980,Morrison2008}.
Yet in its original form optical SR is difficult to observe
due to dephasing dipole-dipole Van der Waals interactions, which suppress a
coherence build-up in atomic ensembles.

In this Letter we show that cooperative emission can occur from the ensemble
of nuclear spins surrounding a quantum emitter such as a self-assembled QD or
an NV center. The interaction of nuclear spin ensemble and optical field is
mediated by the electron spin of the emitter.  Due to the indirect character
of the interaction, the dephasing Van der Waals interactions vanish in this
setting.

We first explain the proposal using the example of an NV center in
diamond. Then, we adapt the model to QDs, which 
promise strong effects due to the large number of involved nuclei.  Despite
the inhomogeneity of the nuclear spin coupling and related dephasing
processes, we predict a SR-like correlation build-up in the nuclear spin
ensemble and a significant intensity burst of several orders of magnitude in
the optical emission profile. 
Finally, we point out the possibility of
observing phase transitions and bistability in the nuclear system.

The superradiant effect is based on the collective hyperfine (HF) interaction
of the electronic spin of the defect (QD or NV) with $N$ initially polarized
proximal nuclear spins. It is dominated by the isotropic contact
term \cite{Schliemann2003,Gali2008} and reads in an external magnetic field
($\hbar=1$):
\begin{equation}
  \label{eq:Ham}
H=\frac{g}{2} (A^+ S^- + A^- S^+) + g A^z S^z + \omega_S S^z.
\end{equation}
Here $S^{\mu}$ and $A^\mu= \sum_{i=1}^N g_i\sigma^{\mu}_i$ ($\mu=+,-,z$)
denote electron and collective nuclear spin operators, respectively. The
coupling coefficients are normalized such that $\sum_i g^2_i=1$ and individual
nuclear spin operators $\sigma^{\mu}_i$ are assumed to be spin-1/2 for
simplicity; $g$ gives the overall HF coupling strength and $\omega_S$ denotes
the electron Zeeman splitting.  We neglect the typically very small nuclear
Zeeman and nuclear dipole-dipole terms.

\begin{figure}[bt]
\centering
\includegraphics[scale=1]{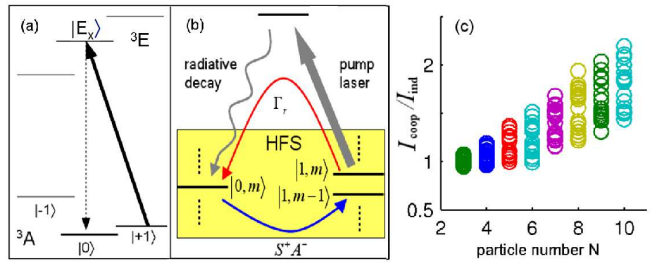}
\caption{(a) Simplified level scheme of NV center with relevant $\Lambda$
  system (cf.~text and \cite{Tamarat2008}); (b) Sketch of relevant processes: electronic ground
  states are coupled by optical pumping and HF flipflops; the states are
  labeled by the $z$-components of the electron and nuclear spin.  (c)
  $I_\mathrm{coop}/I_\mathrm{ind}$ for randomly chosen nuclear environments of
  an NV center The first nuclear shell is taken to be spinless, as due to
  their very strong coupling they would evolve largely independent from the
  ensemble.}
    \label{fig:nv}
\end{figure}

Let us first consider NV centers, in which the effect can be studied in a
clean and relatively small spin environment.  Due to their extraordinary quantum
properties, such as ultra-long decoherence times even at room temperature, NV
centers have attracted wide interest \cite{Jelezko2006} resulting, e.g., in
the demonstration of entanglement and quantum gates between the electron and
proximal nuclear spins \cite{Dutt2007}. Both the NV center's electronic ground
($^3A$) and optically excited states ($^3E$) are spin triplet ($S=1$)
\cite{Jelezko2006}. 
In the absence of a magnetic field, the ground state sublevels
$\ket{m_S=\pm1}$ are split from $\ket{m_S=0}$.  In the following, we assume
that a $B$ field is applied along the NV axis to bring $\ket{m_S=0}$ and
$\ket{m_S=1}$ close to degeneracy\footnote{A field of $\sim100$mT is
  sufficient to compensate the zero-field splitting of $2.88$GHz.}. In this
case $\ket{m_S=-1}$ is off-resonance and can be disregarded. We focus on
low-strain NV centers with well-defined selection rules and assume that it is
optically excited by selectively driving the weakly allowed transition from
$\ket{m_S=1}$ to a state $\ket{E_x}$ in the $^3E$ manifold which decays
primarily into $\ket{m_S=0}$ \cite{Tamarat2008}, see \Figref{fig:nv}(a). 
%
The nuclear spin environment of the NV center consists of proximal $^{13}$C
($I=1/2$) nuclei in the otherwise spinless $^{12}$C matrix, which are
HF-coupled to the electronic spin of the defect center. The interaction is dominated by the
Fermi-contact term such that the coupling is isotropic (to first order) and
described by \Eqref{eq:Ham} ($\omega_S$ here contains both zero field splitting and Zeeman energy). Nevertheless in the simulations conducted below
we included the small anisotropic dipole-dipole terms.

We describe now the coupling of the nuclear spin to the optical field as
depicted in \Figref{fig:nv}(b). It is best understood as a two-step process:
first, strongly driving a dipole-forbidden optical transition of the
$\ket{m_S=1}$ spin state (the allowed transition is far off-resonant) pumps
the electron into $\ket{m_S=0}$. Such Raman spin-flip transitions have been
demonstrated recently \cite{Tamarat2008}. Since the short-lived excited state
is populated negligibly throughout the process, we can eliminate it
from the dynamics using standard techniques and obtain a master equation for
the electron spin decaying with effective rate~$\Gamma_r$:
\begin{equation}
  \label{eq:Meq}
\dot{\rho} = \Gamma_r(S^- \rho S^+ - \frac{1}{2}S^+ S^- \rho -
\frac{1}{2}\rho S^+ S^-) - i [H,\rho],
\end{equation}
each decay being accompanied by a Raman photon. Second, the return to state
$\ket{m_S=1}$, necessary for the next emission, occurs through $H$ via a HF
mediated electron spin flip (and nuclear spin flop). Thus, each Raman photon
indicates a nuclear spin flop and the emission intensity $I(t)$ is
proportional to the change in nuclear polarization.  Starting from a fully
polarized state, SR is due to the increase in the operative HF matrix element
$\left<A^+A^-\right>$.  The scale of the coupling is set by $A:=g\sum_i
g_i$. For a small \emph{relative coupling strength} $\epsilon = A/(2\Delta)$,
where $\Delta:=\left|\Gamma_r/2+i \omega_S\right|$, the electron is
predominantly in its $\ket{m_S=0}$ spin state and we can project
\Eqref{eq:Meq} to the respective subspace.  The reduced master equation for
the nuclear density operator reads
\begin{align}
  \label{eq:Meqelim}
\dot{\rho} = & c_r(A^- \rho A^+ - \frac{1}{2}A^+ A^- \rho -
\frac{1}{2}\rho A^+ A^-)  \\ \nonumber
             & -i c_i[A^+ A^-,\rho]- i g m_S[A^z,\rho],
\end{align}
where $c_r=g^2/(2\Delta)^2\Gamma_r$ and $c_i=g^2/(2\Delta)^2\omega_S$.

As the electron is optically pumped into $\ket{m_S=0}$, the last term -
representing the electron's Knight field - in \Eqref{eq:Meqelim}
vanishes. Assuming resonance ($\omega_S=0$)
the equation closely resembles the
SR master equation which has been discussed extensively in the context of
atomic physics \cite{Gross1982} and thus SR effects might be
expected. However, there is a crucial difference: the \emph{inhomogeneous
  nature} ($g_i\not=\mathrm{const}$) of the operators $A^\mu$: they do not
preserve the collective spin, affecting the relative phase between
nuclei. This could prevent the phased emission necessary for SR
\cite{Gross1982,Leonardi1982,Agarwal1971}. However, as we shall see, SR is
still clearly present in realistic inhomogeneous systems. We take the ratio of
the maximum intensity to the initial intensity (the maximum for independent
spins) $I_\mathrm{coop}/I_\mathrm{ind}$ as our figure of merit in the
following: if this \emph{relative intensity peak height} is $>1$ it indicates
cooperative effects.

To see that this effect can be observed at NV centers, we simulate
\Eqref{eq:Meq} numerically \footnote{\Eqref{eq:Meq} allows for the full
  incorporation of anisotropies.}.  The number $N$ of effectively coupled
nuclei can range from a few to a few hundred, since the concentration of
$^{13}$C can be widely tuned \cite{Collins1988}. The HF constants $g_i$
between the defect and the nearest $\sim 40$ nuclei were derived in
\cite{Gali2008} in an ab-initio calculation. Nuclei outside this shell
($\sim7$\AA) have a coupling strength $gg_i$ weaker than $2\pi \cdot 0.5$MHz
and are not considered here.  The excited state lifetime of the NV center has
been measured as $\tau\approx 13$ns \cite{Lenef1996,Collins1983}. Thus, we
adopt an effective rate $\Gamma_r=2\pi \cdot 10$MHz for the decay from
$\ket{m_S=1}$ to $\ket{m_S=0}$ enabled by driving the Raman transition.  The
intensity enhancements predicted by exact simulations for small, randomly
chosen and initially polarized spin environments are shown in \Figref{fig:nv}(c). In samples of
higher $^{13}$C concentration $N$ can be larger and stronger effects are
expected.

One characteristic feature of SR is the linear $N$-dependence of the
associated effects (already visible in \Figref{fig:nv}).  Since the number of
nuclei to which the electron couples is much larger in a QD than in a
NV center, QDs are particularly attractive candidates for the investigation of
SR.  In the following we study the dynamics of the QD system in different
regimes and we show that strong signatures of SR can be expected in realistic
settings.

Let us consider a self-assembled QD in which a single conduction band electron
is coupled by isotropic Fermi contact interaction to a large number of nuclear
spins.  Optical pumping of the electron is realized by a Raman process,
driving a forbidden transition to a trion state \cite{Akimov}, and including
the HF coupling we again obtain dynamics as sketched in \Figref{fig:nv}(b).
For the optical pumping rate values $\Gamma_r=2\pi \cdot (0.1-1)$GHz are
applicable \cite{Bracker2005,Finkelstein1998}. A comparison with the HF
coupling constants reported for different materials \footnote{The HF coupling
  constants and typical numbers of nuclei ($A/\mu$eV,N) for important QD
  materials are \cite{Akimov,Schliemann2003}: GaAs:(100,$10^4-10^6$),
  CdSe:(10,$10^3$), SiP:(0.1,$10^2$)} shows that for InGaAs and CdSe QDs at
resonance \Eqref{eq:Meqelim} is not valid since the relative coupling strength
$\epsilon \geq 1$. We therefore consider the dynamics of the system under
conditions of a finite electron inversion [using \Eqref{eq:Meq}]. In this
regime, the electron can be seen as a driven and damped two-level system: the
nuclei 'pump' excitations into the electron, which are damped by the
Raman-mediated decay; cooperative behavior manifests in enhanced HF
interaction. This enhancement directly translates into increased electron
inversion $\langle S^+S^-\rangle$ to which the emitted photon rate is
proportional and thus SR from a single QD 
can be expected. 
Let us rephrase this, since SR from a \emph{single} emitter is somewhat
counter-intuitive. Of course, on an optical time scale, anti-bunched single
photons will be emitted at a rate below the optical decay rate. It is, in
fact, typically much slower since the emitter is pumped into the optically
inactive state $\ket{m_S=0}$. SR on time scales $\sim1/\Gamma_r$ consists thus
of lifting this ``spin-blockade'' by HF coupling which becomes increasingly
more efficient as nuclear cooperative effects kick in.
As in the 
homogeneous case \cite{Gross1982} this enhancement is associated with the
transition through nuclear Dicke states $\ket{J,m}, |m|\ll J$. Although $J$ is
not preserved by inhomogeneous $A^\pm$, we can use the Dicke states to
illustrate the dynamics. For instance, due to the large homogeneous component
in $A^-$, its matrix elements show a strong increase $\propto J$ for states
$\ket{J,m}, |m|\ll J$. 

\begin{figure}[thb]
  \begin{center}
    \includegraphics[scale=1]{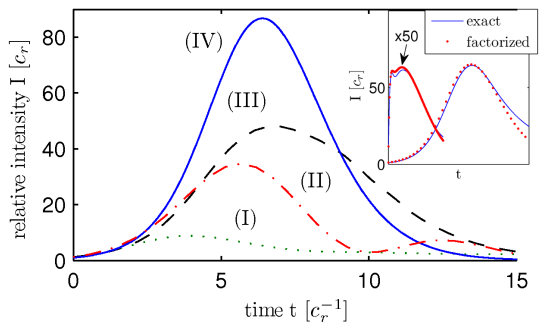}
  \end{center}
    \caption{\label{fig:2} Relative intensity under dynamical Overhauser field
      compensation: $N=21^2$, $\omega_S=A/2$ and $\epsilon=0.3$(I), 0.7(II),
      0.99(III).  (IV) shows the ideal Dicke SR profile \cite{Dicke1954} as a
      reference. Inset: comparison of exact evolution and factorization for
      $N=9$ inhomogeneously coupled spins (left peak, scaled by factor $50$)
      and $N=21^2$ homogeneous spins ($\epsilon=0.7$). Fully independent
      emitters lead to an exponential curve slowly decaying from 1 to zero and
      are therefore not depicted.}
\end{figure}

For large relative coupling strengths $\epsilon \gg 1$ the electron saturates
and superradiant emission is capped by the decay rate $\Gamma_r/2$,
prohibiting the observation of an intensity burst.  In order to avoid this
bottleneck regime we choose a detuning $\omega_S=A/2$ such that $0<
\epsilon=A/\sqrt{\Gamma_r^2 + A^2} \leq 1$.  In this parameter range the early
stage of the evolution - in which the correlation build-up necessary for SR
takes place \cite{Gross1982} - is well described by \Eqref{eq:Meqelim}. The
nuclear phasing is counteracted by the dephasing (inhomogeneous) part of the
Knight term ($\propto g\sqrt{\mathrm{Var}(g_i)}/2$ \cite{Temnov2005}), which
can cause transitions $J\to J-1$.  However, the system evolves in a
\emph{many-body protected manifold} (MPM) \cite{Rey2008}: The term
$\sim[A^+A^-,\rho]$ energetically separates different total nuclear spin-$J$
manifolds. A rough estimate of the ratio between detuning and dephasing shows
a dependence $\propto \epsilon^2$, with proportionality factor $>1$ (diverging
in the homogeneous limit). Thus for values $\epsilon \approx 1$ the
correlation build-up should be largely MPM-protected. We now confirm these
considerations and show by numerical simulation of \Eqref{eq:Meq} that a SR
peaking of several orders of magnitude can be observed in the Raman radiation
from an optically pumped QD, cf.~\Figref{fig:2}.
An exact numerical simulation of the dynamics is not feasible due to the large
number of coupled nuclei and since the dynamics for inhomogeneous coupling
cannot be restricted to a low-dimensional subspace. To obtain
$I(t)\propto\frac{d}{dt}\sum_i\langle \sigma_i^+\sigma_i^-\rangle$ we
therefore use an approximative scheme. By \Eqref{eq:Meq} these expectation
values are related to fourth-order correlation terms involving both the
electron and nuclear spins.  We use a factorization assumption to reduce the
higher-order expressions in terms of the covariance matrix
$\gamma^+_{ij}=\left\langle \sigma^+_i \sigma^-_j\right\rangle$. Following
\cite{AEI93} we apply the bosonic Wick's theorem, incorporating the
fermionic character of same-site nuclear spin operators 
($\left[\sigma^+_i,\sigma^-_i\right]_+ = 1$) and replace, e.g., $\langle
\sigma^+_i \sigma^z_j S^- \rangle \rightarrow (\gamma^+_{jj} - \frac{1}{2})
\langle \sigma^+_i S^-\rangle - \gamma^+_{ij} \langle \sigma^+_j S^- \rangle$.
But the electron spin plays a special role and factorizing it
completely leads to poor results. Therefore we also solve \Eqref{eq:Meq} for
the main higher-order term involving the electron, the ``mediated covariance
matrix'' $\gamma^-_{ij}=\left\langle \sigma^+_i S^z
  \sigma^-_j\right\rangle$. All other higher-order expectation values therein
are factorized under consideration of special symmetries for operators acting
on the same site.

In the regimes accessible to an exact treatment, i.e., the homogeneous case
and for few inhomogeneously coupled particles, the factorization results agree
well with the exact evolution (see inset in \Figref{fig:2}). This shows that
it quantitatively captures the effect of nuclear spin coherences while
allowing a numerical treatment of hundreds of spins.  Finally, in addition to
the constant detuning $\omega_S=A/2$ for the displayed simulations we
compensated the Overhauser field dynamically \footnote{By applying a time
  dependent magnetic or spin-dependent AC Stark field such that
  $\omega_S(t)=g\left<A^z\right>_t$ we ensure that the measured change in
  radiation intensity is due to a cooperative emission effect only.  }.
Furthermore, we assume a Gaussian spatial electron wave function.
\begin{figure}[bht]
\centering
\includegraphics[scale=0.8]{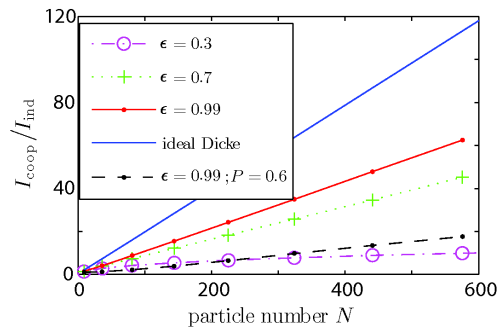}
    \caption{\label{fig:3}$I_\mathrm{coop}/I_\mathrm{ind}$ for different
      values of $\epsilon$ - the Overhauser field is dynamically compensated
      and $\omega_S=A/2$ in all cases - compared to the ideal Dicke
      case. Dashed line corresponds to a partially polarized dot (cf. text).
    }
\end{figure}
The results obtained with these methods are displayed in Figs.~\ref{fig:2} and
\ref{fig:3}.  For $\epsilon\approx1$ the strong MPM protection suppresses
dephasing, leading to pronounced SR signatures: A strong intensity burst,
whose relative height scales $\propto N$ (for large $N$). For
large $\epsilon\approx1$ the relative height is reduced only by half
compared to the ideal Dicke case.  For smaller $\epsilon$ it decreases further
due to increased dephasing. For $\epsilon\leq0.3$, where MPM protection is weak
and the decay process is significantly slowed down ($c_r \propto \epsilon^2$),
even the linear scaling is lost.

From \Figref{fig:3} one extrapolates that for a fully polarized initial state
a huge intensity overhead of several orders of magnitude ($\sim10^3$-$10^4$)
is predicted.  If the initial state is not fully polarized, SR effects are
reduced. However, even when, e.g,. starting from a mixture of symmetric Dicke
states $\ket{J,J}$ with polarization $P=60\%$
\cite{Bracker2005,Maletinsky2008} our simulations predict a strong intensity
peak and (for $N\gg1$) a linear $N$-dependence:
$I_\mathrm{coop}/I_\mathrm{ind} \approx 0.03 N$ ($\epsilon=0.99$), i.e., only
a factor 4 weaker than for full polarization.

Note that for the sake of simplicity we consider $I=1/2$ nuclei in our
simulations. In terms of particle numbers $N$ this is a pessimistic assumption
as typical QD host materials carry a higher spin. We can incorporate this
effect by treating higher spins as $2I$ homogeneously coupled spins $I=1/2$ thus
increasing the effective particle number by the factor $2I$. Most QDs consist
of a few different species of nuclei with strongly varying magnetic moments,
increasing the inhomogeneity of the system. However, in the worst case the
different species evolve independently diminishing the effect by a small
factor corresponding to the number of species. In our simulations the effect
was shown to be much smaller.

We have neglected the dipolar and quadrupolar interaction among the nuclear
spins. The former is always negligible on the time scale considered here
\cite{Schliemann2003}. The latter is absent for nuclear spin $I=1/2$ (NV
centers, CdSe QDs) or strain-free QDs \cite{Bracker2005}. In strongly strained
QDs it can be important \cite{MKI09}, and a term $\sum_i \nu_i (I^{z_i}_i)^2$
must be added to \Eqref{eq:Ham}, where $z_i$ is defined as the main axis of
the local electric-field-gradient tensor.

Having seen that SR can be observed in experimentally accessible nuclear spin
ensembles, let us briefly explore two further applications of this setting:
nuclear spin polarization and phase transitions. We first note that the master
equation \Eqref{eq:Meqelim} describes optical pumping of the nuclear
spins. Its steady states are the eigenstates of $A^z$ which lie in the kernel
of $A^-$, so-called dark states, and include the fully polarized state. Hence
the setting described by Eqs.~(\ref{eq:Meq},\ref{eq:Meqelim}) can be used to
polarize the nuclei \cite{Christ2007}, i.e., to prepare an initial state
required for SR.

Finally, nuclear spin systems may be used to study further cooperative effects
such as phase transitions. It is known \cite{Carmichael1980} that in the
thermodynamic limit an optically driven atomic system with collective decay --
as described by \Eqref{eq:Meqelim} for homogeneous operators and $m_S =
\omega_S =0$ -- can undergo a second-order non-equilibrium phase transition in
the steady state. In our setting an effective driving can be established by a
DC magnetic field $B_x$ perpendicular to the polarization direction. A
semiclassical treatment of the equations of motion deduced from \Eqref{eq:Meq}
predicts a similar phase transition in the combined system of electron and
nuclear spins in certain regimes. Preliminary simulations confirm the validity
of the semiclassical results and also indicate the appearance of related
phenomena like bistability and hysteresis which have recently been observed in
polarization experiments, e.g. \cite{TWR+07}. A detailed analysis of these
topics and an analytical description of the SR dynamics presented here will be
subject of a forthcoming publication \cite{Kessler2010b}.

In conclusion, we have shown that the nuclear spin environment of individual
QDs and NV centers shows superradiant optical emission under suitable optical
pumping conditions. While in NV centers a collective intensity enhancement of
up to 100\% is predicted, the much larger nuclear spin ensembles in QDs could
lead to relative peak heights of several orders of magnitude. This would be
clear evidence of coherent HF dynamics of nuclear spin ensembles in QDs. The
rich physics of SR systems, including bistability and phase transitions could
thus be studied in a long-lived mesoscopic solid-state system.
\begin{acknowledgments}
  We acknowledge support by GIF, the DFG within SFB 631 and the Cluster of Excellence NIM, the NSF, DARPA, and the Packard Foundation.
\end{acknowledgments}


\end{document}